\documentclass{jfm}

\usepackage{amsmath}
\usepackage{amsfonts}
\usepackage{amssymb}
\usepackage{graphicx}
\usepackage{caption}
\usepackage{subcaption}
\usepackage{mathtools}
\usepackage{natbib}
\usepackage[capitalise]{cleveref}

\newcommand{\dint}{\,\textrm{d}}
\newcommand{\mat}[1]{\mathsfbi{#1}}
\renewcommand{\vec}[1]{\textbf{#1}}

\DeclareMathOperator{\J}{J}
\DeclareMathOperator{\R}{R}
\DeclareMathOperator{\JacP}{P}

\title{Modon solutions in an N-layer quasi-geostrophic model}
\author{Matthew N. Crowe\aff{1} and Edward R. Johnson\aff{2}}
\date{}

\shorttitle{N-Layer QG Modons}
\shortauthor{M. N. Crowe \& E. R. Johnson}
\affiliation{\aff{1}School of Mathematics, Statistics and Physics, Newcastle University, Newcastle upon Tyne, NE1 7RU, UK \\
\aff{2}Department of Mathematics, University College London, London, WC1E 6BT, UK}

\begin{document}

\maketitle

\begin{abstract}
Modons, or dipolar vortices, are common and long-lived features of the upper ocean, consisting of a pair of monopolar vortices moving through self-advection. Such structures remain stable over long times and may be important for fluid transport over large distances. Here we present a semi-analytical method for finding fully nonlinear modon solutions in a multi-layer quasi-geostrophic model with arbitrarily many layers. Our approach works by reducing the problem to a multi-parameter linear eigenvalue problem which can be solved using numerical techniques from linear algebra. The method is shown to replicate previous results for one and two-layer models and is applied to a three-layer model to find a solution describing a mid-depth propagating, topographic vortex.
\end{abstract}


\section{Introduction}

Modons, or dipolar vortices, are common coherent structures in the upper ocean \citep{NiZWH20} and are observed to exist for long times and transport fluid over large distances \citep{NycanderI90}. Modons are similarly relevant to atmospheric dynamics, and have been used to model various atmospheric processes such as atmospheric blocking \citep{mcwilliams_1980} and Madden-Julian Oscillation events \citep{rostami_zeitlin_2021}.

Early studies of modon solutions to the two-dimensional Euler equation were carried out independently by Lamb and Chaplygin \citep{Lamb_1932,MeleshkoH94} and similar analytical solutions to the equivalent barotropic, quasi-geostrophic problem were identified by \citet{LarichevR76d}. The long-term behaviour of these structures remains an active area of study, with much recent work devoted to understanding modon breakdown processes such as instabilities \citep{makarov_kizner_2011,FlierlH94,BrionSJ14,JohnsonC21,Davies_et_al_2023,Protas_2024}, energy loss through wave generation \citep{FlierlH94,JohnsonC21}, and viscous decay \citep{FlorVD95,NielsenR97}.

Layered quasi-geostrophic models are commonly used in modelling the ocean and atmosphere. These models consist of two-dimensional layers coupled through a vertical pressure gradient and are generally valid on scales where the Earth's rotation is dominant. Determining modon solutions in such models allows for the study of coherent structures in a more realistic setting than simple 2D dynamics. \citet{Kizner_et_al_2003} thoroughly explore modon solutions in a two-layer model however calculating modon solutions in models with arbitrarily many layers would be mathematically tedious using the same analytical methods.

Here, we extend a method we developed for finding surface QG dipoles \citep{Johnson_Crowe_23,Crowe_Johnson_23} to the layered QG system by analytically reducing the problem to a linear eigenvalue problem which can be solved numerically. We begin by presenting the model and solution method in \cref{sec:setup,sec:sol}, before applying it to some examples in \cref{sec:examples}. Our method is found to be consistent with previous one-layer and two-layer studies and effective at finding new solutions in a three-layer model. Finally, we discuss our method and results in \cref{sec:conc}.

\section{Problem Setup}
\label{sec:setup}

Consider a dipolar vortex (or modon) of radius $a$ moving in the zonal ($x$) direction with speed $U$. In the frame of the moving vortex, the $N$-layer quasi-geostrophic equations are
\begin{equation}
\label{eq:PV_gov}
(\partial_t - U \partial_x) q_i + J(\psi_i,q_i) + \beta_i \partial_x \psi_i = 0, \quad i \in \{1,2,\dots, N\},
\end{equation}
for streamfunction $\psi_i$ and potential vorticity anomaly $q_i$ in each layer. Here $\beta_i$ denotes the background vorticity gradient in the meridional ($y$) direction and may vary by layer to incorporate the effects of bottom topography. The potential vorticities in each layer are given by
\begin{align}
\label{eq:q_def1}
q_1 = &\, \nabla^2 \psi_1 + R_1^{-2} (\psi_2 - \psi_1),\\
\label{eq:q_def2}
q_i = &\, \nabla^2 \psi_i + R_i^{-2} (\psi_{i-1}-2\,\psi_i+\psi_{i+1}), \quad i \in \{2,\dots N-1\},\\
\label{eq:q_def3}
q_N = &\, \nabla^2 \psi_N  + R_N^{-2} (\psi_{N-1}-\psi_N),
\end{align}
where
\begin{equation}
R_i = \frac{\sqrt{g'H_i}}{f},
\end{equation}
is the Rossby radius of deformation in each layer. Here $g'$ denotes the buoyancy difference between consecutive layers and is taken to be the same for all layers for simplicity. Layer depth is denoted by $H_i$ and $f$ denotes the Coriolis parameter. The effects of a background flow can be easily incorporated into this model but will not be considered here.

\cref{eq:PV_gov} can be written as
\begin{equation}
\label{eq:Long}
\partial_t q_i + J(\psi_i+Uy,q_i+\beta_i y) = 0, \quad i \in \{1,2,\dots, N\}
\end{equation}
and we now seek steady, nonlinear modon solutions by letting $\partial_t = 0$. \cref{eq:Long} may be written as
\begin{equation}
\label{eq:long2}
q_i+\beta_i y = F_i(\psi_i+Uy), \quad i \in \{1,2,\dots, N\}
\end{equation}
where $F_i$ is an arbitrary, piece-wise differentiable function of a single variable. To proceed, we take $F_i$ to be a piece-wise linear function
\begin{equation}
\label{eq:Fi}
F_i(z) = \begin{cases}
(\mathcal{K}_i/a^2)\,z, & x^2+y^2 < a^2,\\
(\beta_i/U) \, z, & x^2 + y^2 > a^2,
\end{cases}
\end{equation}
where $a$ denotes the radius of the vortex. The form of the $F_i$ outside the vortex is set by the requirement that vorticity and streamfunction perturbations decay towards infinity. Inside the vortex, we may either choose $\mathcal{K}_i$ to match the exterior solution, i.e. $\mathcal{K}_i = \beta_i a^2/U$, or take $\mathcal{K}_i = -K_i^2$ where $K_i$ is an eigenvalue which will be determined by continuity of $\psi_i$ across the vortex boundary. These cases will be respectively referred to as `passive' and `active' vortex regions. An active region consists of an isolated region of vorticity whereas the vorticity in a passive region is determined only by the advection of the background vorticity, $\beta_i y$, by the vortex flow-field.

\section{Solution Method}
\label{sec:sol}

Using the form of $F_i$ from \cref{eq:Fi}, \cref{eq:long2} can be written as
\begin{align}
\label{eq:sys1}
(\nabla^2-R_1^{-2})\psi_1 + R_1^{-2}\psi_2 + \beta_1 y &= (\mathcal{K}_1/a^2) (\psi_1 + Uy),\\
(\nabla^2 - 2R_i^{-2})\psi_i + R_i^{-2}(\psi_{i-1}+\psi_{i+1}) +\beta_i y &= (\mathcal{K}_i/a^2) (\psi_i + Uy),\\
(\nabla^2 - R_N^{-2})\psi_N + R_N^{-2}\psi_{N-1} + \beta_N y &= (\mathcal{K}_N/a^2) (\psi_N + Uy),
\end{align}
inside the vortex ($x^2+y^2<a^2$) and 
\begin{align}
(\nabla^2-R_1^{-2})\psi_1 + R_1^{-2}\psi_2 &= (\beta_1/U) \psi_1,\\
(\nabla^2 - 2R_i^{-2})\psi_i + R_i^{-2}(\psi_{i-1}+\psi_{i+1})&= (\beta_i/U) \psi_i,\\
\label{eq:sys6}
(\nabla^2 - R_N^{-2})\psi_N + R_N^{-2}\psi_{N-1} &= (\beta_N/U) \psi_N,
\end{align}
outside the vortex ($x^2+y^2>a^2$) for $i \in \{2,\dots N-1\}$.

\subsection{Solution using Hankel transforms}

To proceed, we move to plane polar coordinates, $(x,y) = (r\cos\theta,r\sin\theta)$, and represent $\psi_i$ using the Hankel tranform
\begin{equation}
\label{eq:psi_hat}
\psi_i = Ua\,\sin\theta \int_0^\infty \hat{\psi}_i(\xi) \J_1(s\xi)\, \xi \dint \xi,
\end{equation}
where $s = r/a$ and $\J_1$ denotes the Bessel function of the first kind of order $1$. The angular dependence in \cref{eq:psi_hat} is chosen to match that of a typical dipolar vortex. Substituting \cref{eq:psi_hat} into Eqs. (\ref{eq:sys1}) to (\ref{eq:sys6}) gives
\begin{align}
\label{eq:int1}
\int_0^\infty \left[(\xi^2+\lambda_1^2 +\mathcal{K}_1)\hat{\psi}_1 - \lambda_1^2 \hat{\psi}_2\right] \J_1(s\xi)\, \xi \dint \xi  &= (\mu_1-\mathcal{K}_1) s,\\
\int_0^\infty \left[(\xi^2+2\lambda_i^2 +\mathcal{K}_i)\hat{\psi}_i - \lambda_i^2 (\hat{\psi}_{i+1} + \hat{\psi}_{i-1})\right] \J_1(s\xi)\, \xi \dint \xi  &= (\mu_i-\mathcal{K}_i) s,\\
\int_0^\infty \left[(\xi^2+\lambda_N^2 +\mathcal{K}_N)\hat{\psi}_N - \lambda_N^2 \hat{\psi}_{N-1}\right] \J_1(s\xi)\, \xi \dint \xi  &= (\mu_N-\mathcal{K}_N) s,
\end{align}
inside the vortex ($s<1$) and
\begin{align}
\int_0^\infty \left[(\xi^2+\lambda_1^2 +\mu_1)\hat{\psi}_1 - \lambda_1^2 \hat{\psi}_2\right] \J_1(s\xi)\, \xi \dint \xi  &= 0,\\
\int_0^\infty \left[(\xi^2+2\lambda_i^2 +\mu_i)\hat{\psi}_i - \lambda_i^2 (\hat{\psi}_{i+1} + \hat{\psi}_{i-1})\right] \J_1(s\xi)\, \xi \dint \xi  &= 0,\\
\label{eq:int6}
\int_0^\infty \left[(\xi^2+\lambda_N^2 +\mu_N)\hat{\psi}_N - \lambda_N^2 \hat{\psi}_{N-1}\right] \J_1(s\xi)\, \xi \dint \xi  &= 0,
\end{align}
outside the vortex ($s>1$) for $i \in \{2,\dots N-1\}$. Here $\mu_i = \beta_i a^2/U$ and $\lambda_i = a/R_i$ respectively denote the nondimensional background vorticity gradient and the ratio of the vortex radius to the Rossby radius in each layer.

We now define the matrix function
\begin{equation}
\mat{K}(\xi) = \begin{pmatrix}
\xi^2+\lambda_1^2 & -\lambda_1^2 & 0 & \dots & 0 & 0\\
-\lambda_2^2 & \xi^2+2\lambda_2^2 & -\lambda_2^2 & \dots & 0 & 0 \\
\vdots & \vdots & \vdots & \ddots & \vdots & \vdots \\
0 & 0 & 0 & \dots & -\lambda_N^2 & \xi^2+\lambda_N^2 
\end{pmatrix},
\end{equation}
and diagonal matrix
\begin{equation}
\mat{D}(\vec{c}) = \begin{pmatrix}
c_1 & 0 & \dots & 0 \\
0 & c_2 & \dots & 0 \\
\vdots & \vdots & \ddots & \vdots \\
0 & 0 & \dots & c_N 
\end{pmatrix},
\end{equation}
for some vector $\vec{c} = (c_1, c_2, \dots c_N)^T$. We also introduce the vectors $\boldsymbol{\mu} = (\mu_1, \mu_2, \dots, \mu_N)^T$, $\boldsymbol{\mathcal{K}} = (\mathcal{K}_1, \mathcal{K}_2, \dots, \mathcal{K}_N)^T$ and $\hat{\boldsymbol{\psi}} = (\hat{\psi}_1, \hat{\psi}_2, \dots, \hat{\psi}_N)^T$ so Eqs. (\ref{eq:int1}) to (\ref{eq:int6}) can be written concisely as
\begin{align}
\label{eq:psi_vec1}
\int_0^\infty \left[\mat{K}(\xi)+\mat{D}(\boldsymbol{\mathcal{K}}) \right]\hat{\boldsymbol{\psi}}(\xi)\, \J_1(s\xi)\, \xi \dint \xi &= (\boldsymbol{\mu} - \boldsymbol{\mathcal{K}})s, & s < 1,\\
\label{eq:psi_vec2}
\int_0^\infty \left[ \mat{K}(\xi)+\mat{D}(\boldsymbol{\mu})\right]\hat{\boldsymbol{\psi}}(\xi)\, \J_1(s\xi)\, \xi \dint \xi &= 0, & s > 1.
\end{align}

To proceed, we follow the approach of \citet{Johnson_Crowe_23} and let
\begin{equation}
\label{eq:A_def}
\vec{A}(\xi) = \left[ \mat{K}(\xi)+\mat{D}(\boldsymbol{\mu})\right]\hat{\boldsymbol{\psi}}(\xi)\,\xi,
\end{equation}
where $\vec{A}$ is expanded in terms of Bessel functions as
\begin{equation}
\label{eq:A_exp}
\vec{A}(\xi) = \sum_{j = 0}^\infty \vec{a}_j \J_{2j+2}(\xi),
\end{equation}
where the $\vec{a}_j$ are vectors of expansion coefficients. This expansion allows us to exploit the integral relation
\begin{equation}
\label{eq:int_rel1}
\int_0^\infty \J_{2j+2}(\xi) \J_1(s\xi) \dint \xi = \begin{cases}
\R_j(s) & \textrm{for} \quad s < 1,\\ 0 & \textrm{for} \quad s > 1,
\end{cases}
\end{equation}
and its inverse result
\begin{equation}
\J_{2k+2}(\xi)/\xi = \int_0^1 s \R_k(s) \J_1(s\xi) \dint s,
\end{equation}
where
\begin{equation}
\R_n(s) = (-1)^n s \JacP^{(0,1)}_n(2s^2-1),
\end{equation}
denotes the Zernike radial function, a set of polynomials of degree $2n+1$ \citep{BornW19} which are orthogonal over $s\in[0,1] $ with weight $s$, and $\JacP_n^{(\alpha,\beta)}$ denotes the Jacobi polynomial. From \cref{eq:int_rel1} for $s>1$, we observe that \cref{eq:psi_vec2} is automatically satisfied by our expansion of $\vec{A}(\xi)$ in \cref{eq:A_exp} for all choices of $\vec{a}_j$.

\cref{eq:psi_vec1} may now be written as
\begin{equation}
\label{eq:int_rel3}
\sum_{j = 0}^\infty\left[\int_0^\infty \left[\mat{K}(\xi)+\mat{D}(\boldsymbol{\mathcal{K}}) \right]\left[ \mat{K}(\xi)+\mat{D}(\boldsymbol{\mu})\right]^{-1} \J_{2j+2}(\xi) \J_1(s\xi) \dint \xi\right]\vec{a}_j  = (\boldsymbol{\mu} - \boldsymbol{\mathcal{K}})s,
\end{equation}
for $s < 1$. We now multiply \cref{eq:int_rel3} by $s\R_k(s)$ and integrate over $s\in [0,1]$ to get
\begin{equation}
\label{eq:int_rel4}
\sum_{j = 0}^\infty \left[\int_0^\infty \left[\mat{K}(\xi)+\mat{D}(\boldsymbol{\mathcal{K}}) \right]\left[ \mat{K}(\xi)+\mat{D}(\boldsymbol{\mu})\right]^{-1} \xi^{-1} \J_{2j+2}(\xi) \J_{2k+2}(\xi) \dint \xi \right]\vec{a}_j  = \frac{\delta_{k0}}{4}(\boldsymbol{\mu} - \boldsymbol{\mathcal{K}}),
\end{equation}
where $k \in \{0,1,2,\dots\}$. Defining
\begin{align}
\mat{A}_{kj} &= \int_0^\infty \mat{K}(\xi) \left[ \mat{K}(\xi) +\mat{D}(\boldsymbol{\mu})\right]^{-1} \xi^{-1} \J_{2j+2}(\xi) \J_{2k+2}(\xi) \dint \xi, \\
\mat{B}_{kj} &= \int_0^\infty \left[ \mat{K}(\xi) +\mat{D}(\boldsymbol{\mu})\right]^{-1} \xi^{-1} \J_{2j+2}(\xi) \J_{2k+2}(\xi) \dint \xi,\\
c_k &= \frac{\delta_{k0}}{4},
\end{align}
allows us to write \cref{eq:int_rel4} as
\begin{equation}
\label{eq:sol1}
\sum_{j=0}^\infty \left[ \mat{A}_{kj} + \mat{D}(\boldsymbol{\mathcal{K}})\mat{B}_{kj}\right] \vec{a}_j = (\boldsymbol{\mu}-\boldsymbol{\mathcal{K}})c_k.
\end{equation}
We may now truncate this sum after a finite number of terms, say $M$, so $j,k \in \{0,1,2,\dots, M-1\}$. The resulting problem is an $NM \times NM$ inhomogeneous, eigenvalue problem with up to $N$ eigenvalues. Up to $N$ additional equations are needed to solve this system and may be obtained by the requirement that the vortex boundary is a streamline in the vortex frame, $\psi_i +Uy = 0$ on $s = 1$. This condition gives $N$ equations in the form of a single equation for the coefficient vectors $\vec{a}_j$:
\begin{equation}
\label{eq:cond}
\int_0^\infty \vec{A}(\xi) \J_1(\xi)\, \xi \dint \xi = \sum_{j = 0}^\infty \vec{a}_j \!\int_0^\infty \J_{2j+2}(\xi) \J_1(\xi) \dint \xi = 0 \!\quad \!\implies \!\quad \sum_{j = 0}^\infty (-1)^j \vec{a}_j = 0,
\end{equation}
which may similarly be truncated after $M$ terms.

We note that modon solutions do not exist if there's a singularity of $[\mat{K}(\xi) +\mat{D}(\boldsymbol{\mu})]^{-1}$ for $\xi \in (0,\infty)$ as the integrals $\mat{A}_{kj}$ and $\mat{B}_{jk}$ do not converge. These singularities corresponds to the excitation of linear waves in the layered model. In the 2D case, the condition for the existence of singularities is equivalent to the conditions discussed in \citet{Kizner_et_al_2003}.

\subsection{The truncated eigenvalue problem}

Our truncated eigenvalue problem from \cref{eq:sol1,eq:cond} may be written as
\begin{equation}
\label{eq:fullEVP}
\left[ \mat{A} + \sum_{i = 1}^N \mathcal{K}_i \mat{B}_i\right]\vec{a} = \sum_{i=1}^N (\mu_i-\mathcal{K}_i)\vec{c}_i, \quad \textrm{s.t.}\quad \vec{d}_i^T\vec{a} = 0 \quad \textrm{for} \quad i\in\{1,\dots,N\},
\end{equation}
for
\begin{equation}
\mat{A} = \begin{pmatrix}
\mat{A}_{0,0} &\dots & \mat{A}_{0,M-1} \\
\vdots & \ddots & \vdots \\
\mat{A}_{M-1,0} &\dots & \dots \mat{A}_{M-1,M-1}
\end{pmatrix},\quad
\mat{B} = \begin{pmatrix}
\mat{B}_{0,0} &\dots & \mat{B}_{0,M-1} \\
\vdots & \ddots & \vdots \\
\mat{B}_{M-1,0} &\dots & \dots \mat{B}_{M-1,M-1}
\end{pmatrix},
\end{equation}
$\vec{a} = (\vec{a}_0^T,\vec{a}_1^T,\dots,\vec{a}_{M-1}^T)^T$,  $\vec{c} = (\boldsymbol{1},\boldsymbol{0},\dots,\boldsymbol{0})^T$ and $\vec{d} = (\boldsymbol{1},-\boldsymbol{1},\dots,(-1)^{M-1} \boldsymbol{1})^T$ where $\vec{1}$ and $\vec{0}$ denote length $N$ row vectors of ones and zeros respectively. The indexed quantities $\mat{B}_i$ and $\vec{c}_i$ are given by $\mat{B}$ and $\vec{c}$ respectively where all rows with a row index, $n \neq i \pmod{N}$, are set to zero.

\cref{eq:fullEVP} is a multi-parameter eigenvalue problem \citep{MPEVP_book} and may be solved using a range of methods (see \cref{sec:appA}). For active layers, we must solve for the eigenvalue $\mathcal{K}_i = -K_i^2$ while for passive layers we let $\mathcal{K}_i = \mu_i$ so the term $\mathcal{K}_i\mat{B}_i$ may be merged into $\mat{A}$ and the right-hand side term $(\mu_i-\mathcal{K}_i)\vec{c}_i$ vanishes. Additional conditions of the form $\vec{d}_i^T\vec{a} = 0$ are required for active layers but not for passive layers where continuity is automatically enforced by the fact that the components of $\vec{a}$ corresponding to a passive layer are zero. Mathematically, this is equivalent to the requirement to include an additional equation each time a new (unknown) eigenvalue is introduced. There will be a discrete spectrum of Eigenvalues for each layer, corresponding to different modes in the radial direction. However, modon studies typically focus on the first order radial mode which has the smallest value of $K_i$ and is generally the most stable. For multi-layer models, modon solutions may have different radial mode numbers in each layer.

\subsection{Determining streamfunctions and potential vorticies}

Once the $K_i$ and $\vec{a}_j$ are calculated, we may determine the streamfunctions, $\psi_i$, by inverting \cref{eq:A_def} for $\hat{\boldsymbol{\psi}}_i(\xi)\,\xi$ and evaluating the Hankel transform in \cref{eq:psi_hat} directly. Alternatively, using \cref{eq:A_def,eq:A_exp,eq:int_rel1} we can show that
\begin{equation}
\label{eq:psi_sol}
\boldsymbol{\Delta}_N(\boldsymbol{\beta})\, \boldsymbol{\psi} = -\frac{U}{a} \sin\theta \sum_{j=0}^{M-1} \vec{a}_j \R_j(r/a),
\end{equation}
where $\boldsymbol{\psi} = (\psi_1, \psi_2,\dots, \psi_N)^T$ and
\begin{equation}
\boldsymbol{\Delta}_N(\boldsymbol{\beta}) = \begin{pmatrix}
\nabla^2 - \frac{1}{R_1^2}-\frac{\beta_1}{U} & \frac{1}{R_1^2} & 0 & \dots & 0 & 0\\
\frac{1}{R_2^2} & \nabla^2 - \frac{2}{R_2^2}-\frac{\beta_2}{U} & \frac{1}{R_2^2} & \dots & 0 & 0\\
\vdots & \vdots & \vdots & \ddots & \vdots & \vdots \\
0 & 0 & 0 & \dots & \frac{1}{R_N^2} & \nabla^2 - \frac{1}{R_N^2} - \frac{\beta_N}{U}
\end{pmatrix},
\end{equation}
where $\boldsymbol{\beta} = (\beta_1, \beta_2,\dots,\beta_N)^T$. \cref{eq:psi_sol} may be easily inverted using discrete Fourier transforms so this method is often easier numerically than inverting the Hankel transforms. From \cref{eq:q_def1,eq:q_def2,eq:q_def3}, potential vorticity anomalies may be similarly calculated using
$\vec{q} = \Delta_N(\boldsymbol{0})\, \boldsymbol{\psi}$ where $\vec{q} = (q_1,q_2,\dots,q_N)^T$.

\section{Examples}
\label{sec:examples}

We now present three examples to demonstrate that this method gives consistent results with previous studies and can be used to find new modon solutions in layered QG models. A Matlab script to solve for the general N-layer problem is included as supplementary material.

\subsection{The Larichev and Reznik dipole}

\begin{figure}
\centering
\begin{subfigure}{0.49\textwidth}
\includegraphics[width=\textwidth,trim={0 0 0 0},clip]{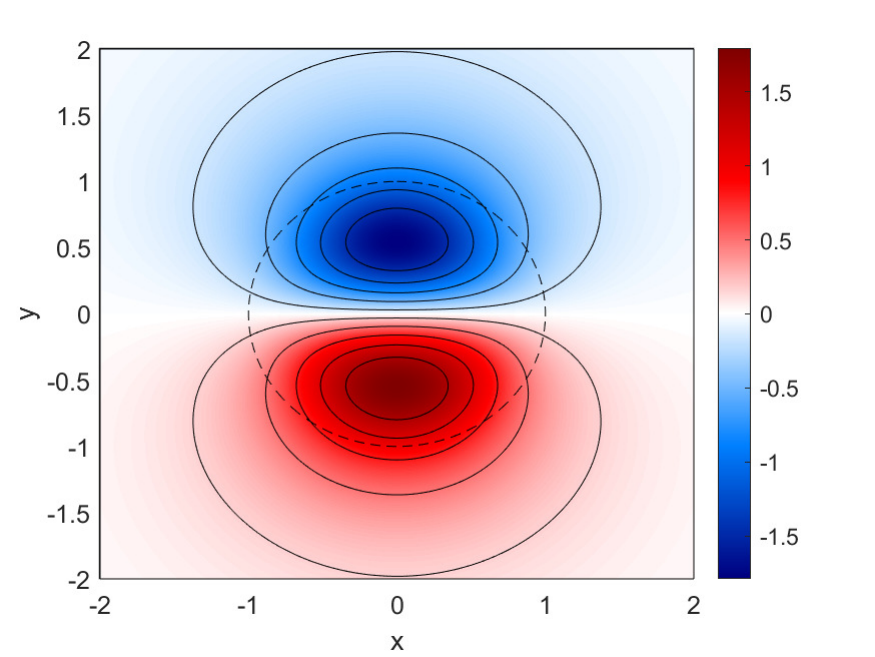}
\caption{}
\end{subfigure}
\begin{subfigure}{0.49\textwidth}
\includegraphics[width=\textwidth,trim={0 0 0 0},clip]{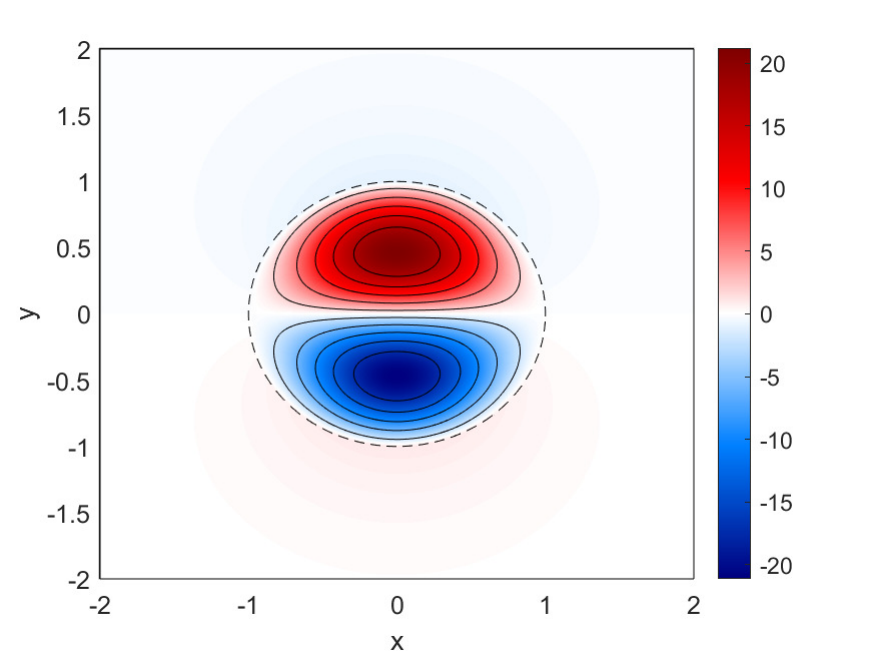}
\caption{}
\end{subfigure}
\caption{A 1-layer modon solution with $(U,a,R_1,\beta_1) = (1,1,1,1)$. We show (a) $\psi_1$ and (b) $q_1$. This solution corresponds to an eigenvalue of $K_1 \approx 4.108$.}
\label{fig:mod1}
\end{figure}

Consider a 1-layer model and choose parameters $(U,a,R_1,\beta_1) = 1$ and $\boldsymbol{\mathcal{K}}_1 = -K_1^2$. Modon solutions may be found using the method of \citet{LarichevR76d} which combines a Bessel function solution with root finding to determine $K_1$. Using our approach, we construct a linear problem of the form
\begin{equation}
\left[\mat{A} - K_1^2\mat{B}\right]\,\vec{a} = (\mu_1+K_1^2)\,\vec{c},\quad \textrm{s.t.} \quad \vec{d}^T\vec{a} = 0,
\end{equation}
using \cref{eq:sol1,eq:cond}. This system may be solved using the method outlined in \cref{sec:appA}. We determine a value of $K = 4.10787...$ which matches the value calculated via root finding to seven significant figures using only $M = 12$ terms. Plots of the streamfunction $\psi_1$ and potential vorticity anomaly $q_1$ are given in \cref{fig:mod1}. Higher order radial modes and solutions for different parameters, including the case of $R_1 = \infty$, may be easily calculated using the same method.

\subsection{Beta-plane baroclinic topographic modons}

\begin{figure}
\centering
\begin{subfigure}{0.49\textwidth}
\includegraphics[width=\textwidth,trim={0 0 0 0},clip]{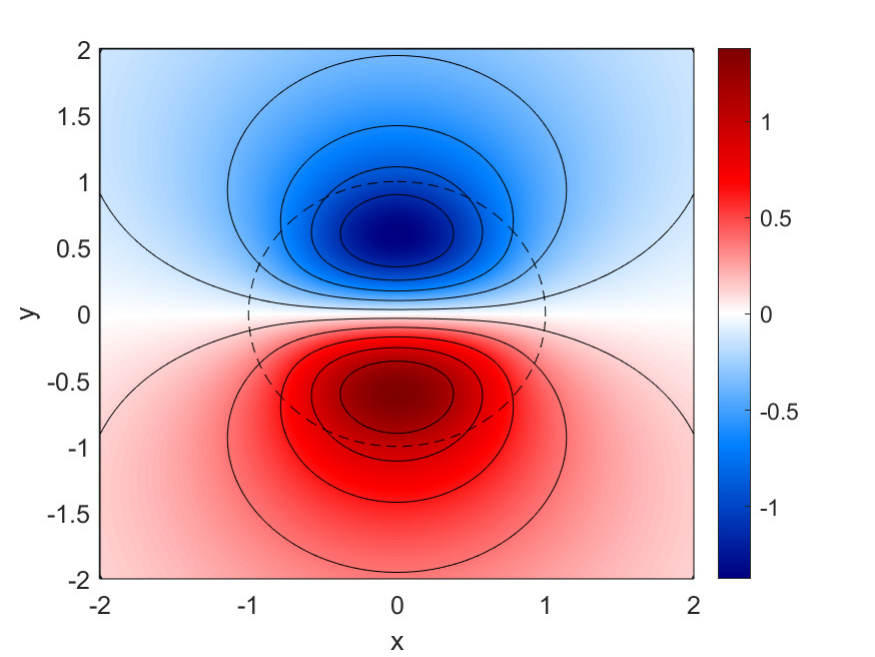}
\caption{}
\end{subfigure}
\begin{subfigure}{0.49\textwidth}
\includegraphics[width=\textwidth,trim={0 0 0 0},clip]{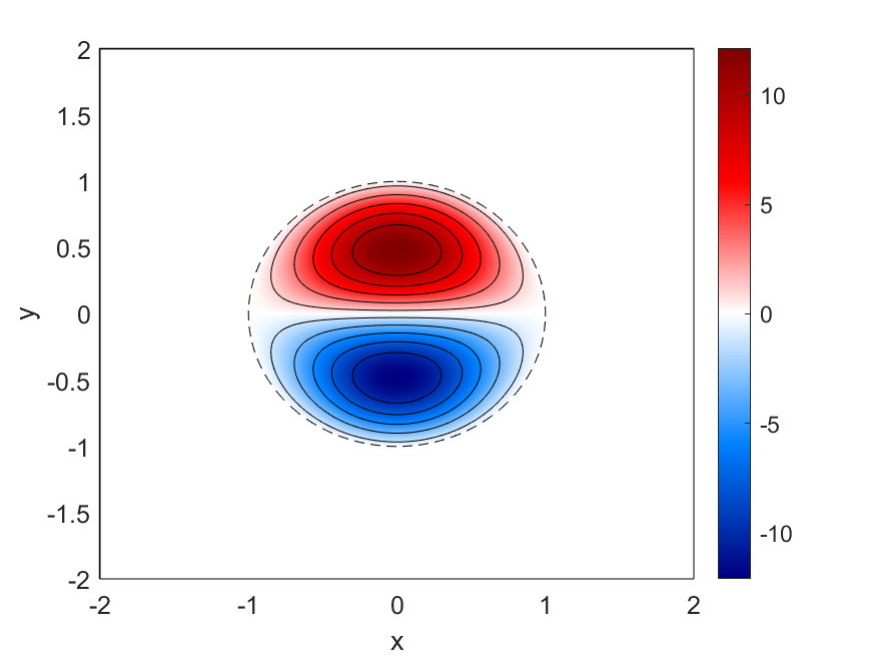}
\caption{}
\end{subfigure}
\\
\begin{subfigure}{0.49\textwidth}
\includegraphics[width=\textwidth,trim={0 0 0 0},clip]{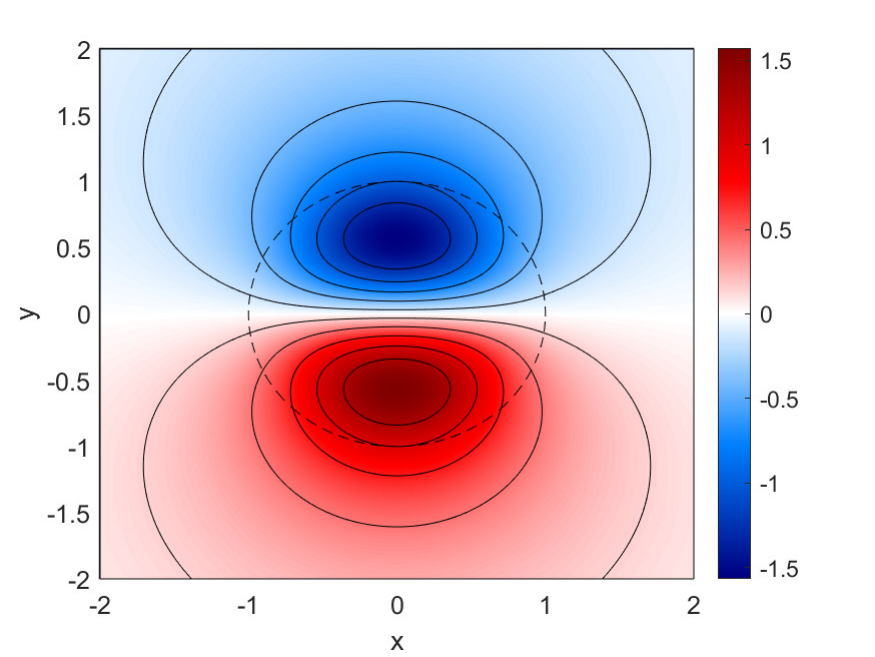}
\caption{}
\end{subfigure}
\begin{subfigure}{0.49\textwidth}
\includegraphics[width=\textwidth,trim={0 0 0 0},clip]{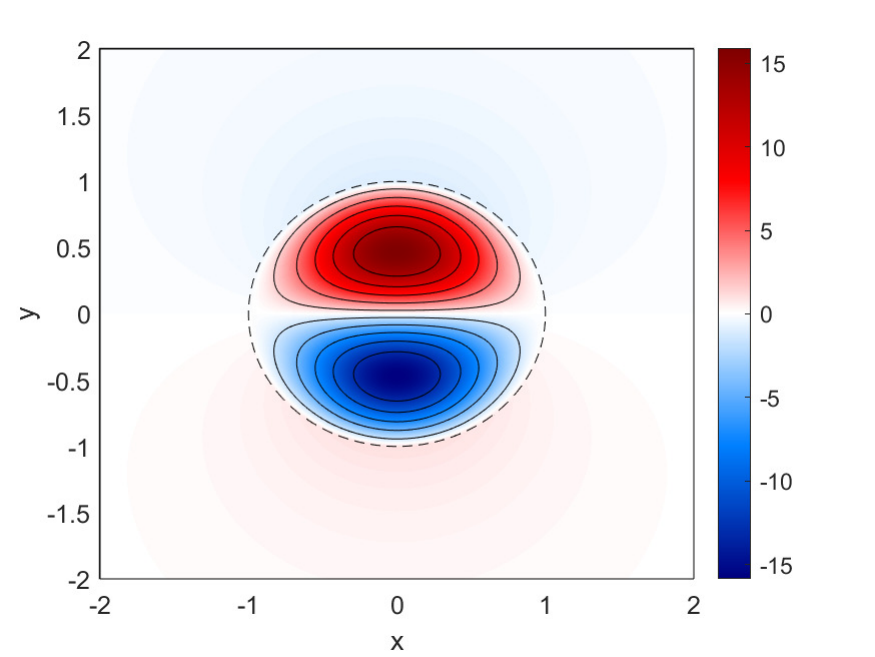}
\caption{}
\end{subfigure}
\caption{A 2-layer modon solution with $(U,a,R_1,R_2,\beta_1,\beta_2) = (1,1,1,1,0,1)$. We show (a) $\psi_1$, (b) $q_1$, (c) $\psi_2$, and (d) $q_2$. This solution corresponds to eigenvalues $(K_1,K_2) \approx (3.800, 3.950)$.}
\label{fig:mod2}
\end{figure}

\citet{Kizner_et_al_2003} provide an extensive discussion of baroclinic beta-plane topographic modons using a two layer quasi-geostrophic model. Here we present a two layer solution with $(U,a,R_1,R_2,\beta_1,\beta_2) = (1,1,1,1,0,1)$ and $(\mathcal{K}_1,\mathcal{K}_2) = -(K_1^2,K_2^2)$ corresponding to two active layers. Using our approach, this problem is straightforwardly reduced to solving the two-parameter eigenvalue problem
\begin{equation}
\left[ \mat{A} - K_1^2 \mat{B}_1 - K_2^2 \mat{B}_2\right]\vec{a} = (\mu_1+K_1^2)\vec{c}_1 + (\mu_2+K_2^2)\vec{c}_2,\quad\textrm{s.t.}\quad \vec{d}_1^T\vec{a} = \vec{d}_2^T\vec{a} = 0,
\end{equation}
which may be approached using the methods outlined in \cref{sec:appA}. Our 2-layer modon solutions are shown in \cref{fig:mod2}. Solutions with one passive layer---referred to as `modons with one interior domain' by \citet{Kizner_et_al_2003}---may be obtained by solving the same problem using $K_2^2 = -\mu_2$ and neglecting the condition $\vec{d}_2^T\vec{a} = 0$. This demonstrates a major advantage of our approach; different vortex regimes can be considered using the same problem setup. By contrast, when using the Bessel function approach of \citet{Kizner_et_al_2003}, care is needed to ensure that the correct Bessel function is used depending on the sign of $\mathcal{K}_i$. As such, switching between active and passive layers is less straightforward as it requires both a solution with a different functional form, and new matching conditions across the vortex boundary.

\subsection{A mid-depth topographic modon}

\begin{figure}
\centering
\begin{subfigure}{0.49\textwidth}
\includegraphics[width=\textwidth,trim={0 0 0 0},clip]{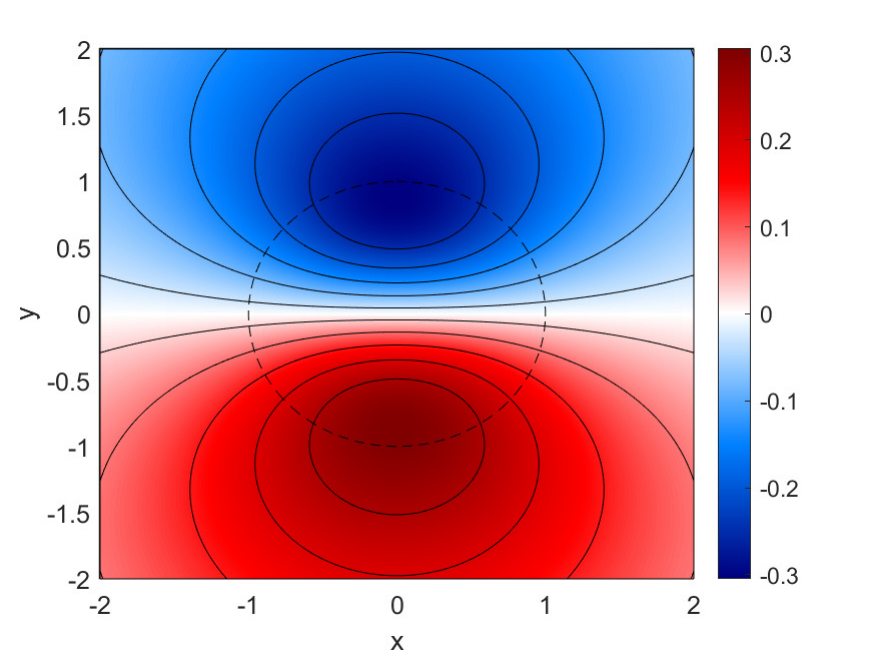}
\caption{}
\end{subfigure}
\begin{subfigure}{0.49\textwidth}
\includegraphics[width=\textwidth,trim={0 0 0 0},clip]{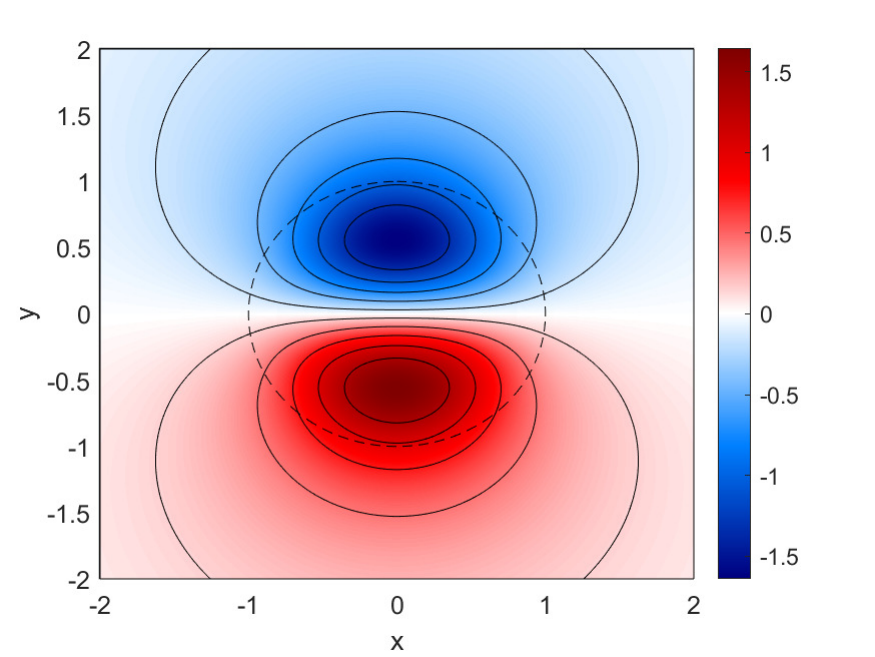}
\caption{}
\end{subfigure}
\\
\begin{subfigure}{0.49\textwidth}
\includegraphics[width=\textwidth,trim={0 0 0 0},clip]{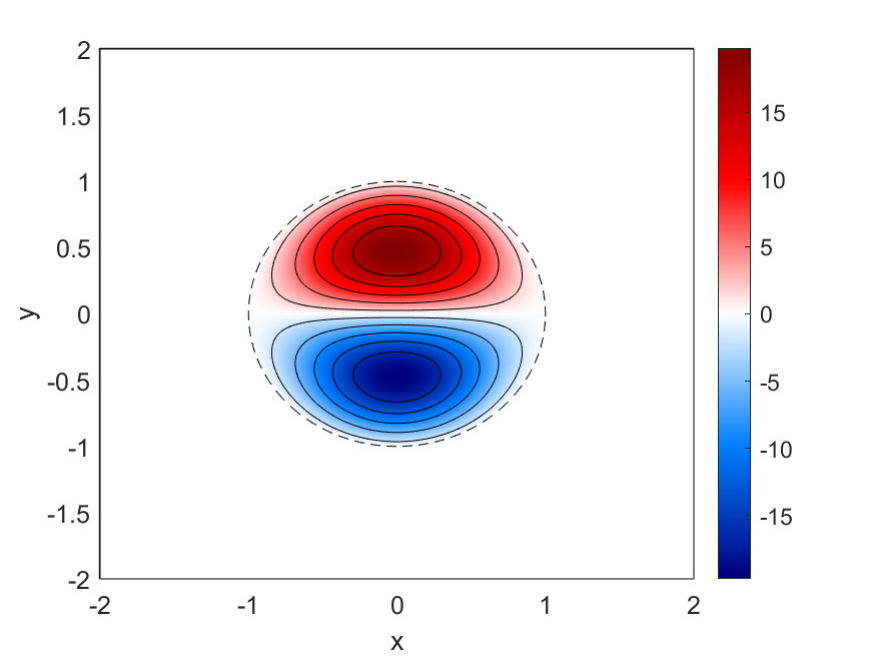}
\caption{}
\end{subfigure}
\begin{subfigure}{0.49\textwidth}
\includegraphics[width=\textwidth,trim={0 0 0 0},clip]{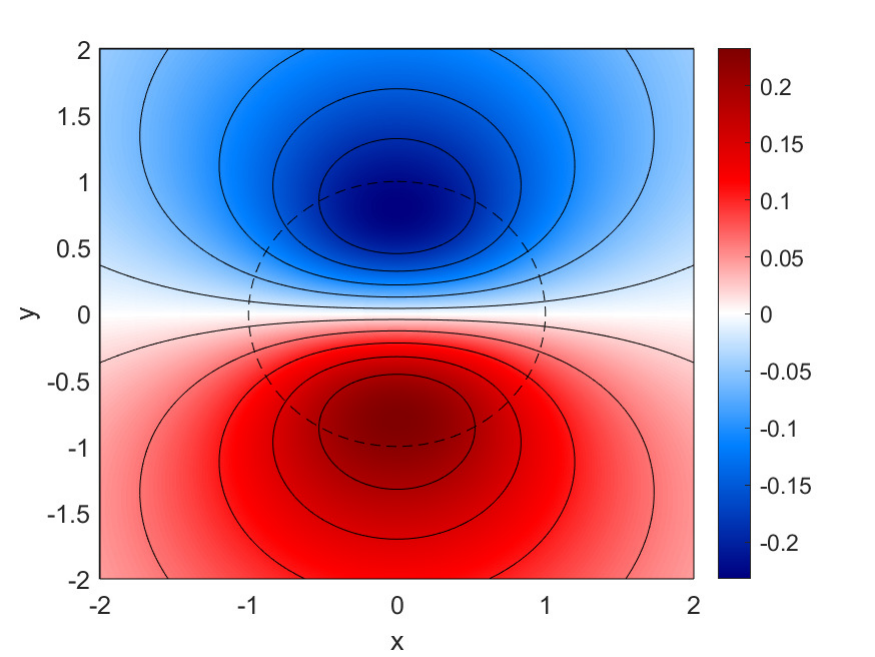}
\caption{}
\end{subfigure}
\caption{A 3-layer mid-depth vortex solution with $(U,a,R_1,R_2,R_3,\beta_1,\beta_2,\beta_3) = (1,1,1,1,1,0,0,1)$. We show (a) $\psi_1$, (b) $\psi_2$, (c) $q_2$, and (d) $\psi_3$. In this case, $q_1 = 0$ and $q_3 = \psi_3$. Layers 1 and 3 are passive and layer 2 has eigenvalue $K_2 \approx 4.1835$.}
\label{fig:mod3}
\end{figure}

To demonstrate the flexibility of our approach, we consider a mid-depth vortex in a 3-layer model. We assume that the top and bottom layers are passive and consider an active middle layer. A topographic slope with gradient in the $y$ direction is modelled by taking $\beta_1 = 0, \beta_2 = 0$ in the top two layers and $\beta_3 = 1$ in the bottom layer. For simplicity we take $R_i = 1$ for $i \in \{1,2,3\}$, $a = 1$ and $U = 1$. This problem reduces to the eigenvalue problem
\begin{equation}
\left[ \mat{A} +\mu_1 \mat{B}_1 - K_2^2 \mat{B}_2 + \mu_3\mat{B}_3\right]\vec{a} = (\mu_2+K_2^2)\vec{c}_2, \quad\textrm{s.t.}\quad \vec{d}_2^T\vec{a} = 0,
\end{equation}
which can be straightforwardly solved using standard methods (see \cref{sec:appA}). The solution consists of an isolated region of strong vorticity in the middle layer, with velocity fields in all three layers. Weak vorticity is generated in the bottom layer due to the advection of fluid in the up-slope ($y$) direction while no vorticity anomaly exists in the surface layer. \cref{fig:mod3} shows $\psi_i$ for $i \in \{1,2,3\}$ and $q_2$ for this solution, the fields $q_1 = 0$ and $q_3 = \psi_3$ are not shown.

\section{Discussion and Conclusions}
\label{sec:conc}

We have presented a semi-analytical method for finding modon, or dipolar vortex, solutions in a layered quasi-geostrophic model. While the system is, in principle, solvable using combinations of Bessel functions inside and outside the vortex region, the matching conditions required to determine all the coefficients can become extremely complicated for multiple layers. Further, these matching conditions must be solved numerically with solutions depending on the vortex parameters in a complicated, nonlinear way. As such, any additional insight obtained from having a closed form solution is limited. By contrast, our procedural method can be straightforwardly used to convert a model with arbitrarily many layers into a multi-parameter eigenvalue problem \citep{MPEVP_book} which can be solved numerically.

Additional advantages of our method include the ability to solve for any combination of active and passive layers without having to recalculate any terms in the linear system or change the form of the solution. Though we have not examined this in detail, we expect conditions for the existence of steady modon solutions---corresponding to a lack of linear wave excitation---to follow from examining the singularities of the quantity $[\mat{K}+\mat{D}(\boldsymbol{\mu})]^{-1}$. Finally, cases which would have to be treated separately using the standard analytical approach, such as an infinite Rossby radius, require no special treatment here.

We suggest that methods combining Hankel transforms and Zernike radial functions may be relevant to a range of physical problems where important dynamics occur in an isolated circular region. These methods are not restricted to variants of the Laplace and Helmholtz equations, as shown for the Dirichlet-to-Neumann operators of \citet{Crowe_Johnson_23,Johnson_Crowe_23}. Additional effects---such as variations in density difference between layers---could be easily included into our model. It may also be possible to include the effects of layer-dependent background flow or extend the method to a three-dimensional quasi-geostrophic model.

\noindent{\bf Code Availability\bf{.}} Matlab scripts implementing this method for the N-layer case and demonstrating the three examples considered are included as supplementary material.

\noindent{\bf Declaration of Interests\bf{.}} The authors report no conflict of interest.

\appendix

\section{Notes on Inhomogeneous Eigenvalue Problems}
\label{sec:appA}

Consider the problem
\begin{equation}
\label{eq:EVP}
(\mat{A} - \lambda \mat{B})\, \vec{a} = \vec{c}_0 + \lambda \vec{c}_1, \quad \textrm{such that} \quad \vec{d}^T\vec{a} = 0.
\end{equation}
Assuming that $\mat{A}$ is non-singular, we can left-multiply \cref{eq:EVP} by $\vec{d}^T\mat{A}^{-1}$ and multiply the resulting scalar by $\vec{c}_0+\lambda\vec{c}_1$ to get
\begin{equation}
-\frac{\lambda(\vec{c}_0\vec{d}^T+\lambda\vec{c}_1\vec{d}^T) \mat{A}^{-1}\mat{B}}{(\vec{d}^T\mat{A}^{-1}\vec{c}_0) + \lambda( \vec{d}^T\mat{A}^{-1}\vec{c}_1)}\,\vec{a} = \vec{c}_0+\lambda\vec{c}_1,
\end{equation}
which can be used to eliminate the inhomogeneity from \cref{eq:EVP}. The resulting homogeneous problem is the quadratic eigenvalue problem
\begin{multline}
\label{eq:quad_EVP}
\left[ (\vec{d}^T \mat{A}^{-1} \vec{c}_0) \mat{A} + \lambda \left( (\vec{d}^T\mat{A}^{-1} \vec{c}_1)\mat{A} - (\vec{d}^T\mat{A}^{-1}\vec{c}_0)\mat{B} - (\vec{c}_0 \vec{d}^T)\mat{A}^{-1}\mat{B} \right) \right. \\ \left. - \lambda^2  \left( (\vec{d}^T\mat{A}^{-1} \vec{c}_1)\mat{B} + (\vec{c}_1\vec{d}^T)\mat{A}^{-1}\mat{B} \right) \right] \vec{a} = 0.
\end{multline}
\cref{eq:quad_EVP} may be put into canonical form and solved as a linear eigenvalue problem for $\lambda$. The corresponding eigenvector, $\vec{a}$, should be directly calculated from \cref{eq:EVP} as the eigenvectors from \cref{eq:quad_EVP} will not necessarily have the correct magnitude since the quadratic problem is linear in $\vec{a}$.


Consider the general problem
\begin{equation}
\left[ \mat{A} - \sum_{i = 1}^N \lambda_i \mat{B}_i \right] \vec{a} = \vec{c}_0 + \sum_{i=1}^N \lambda_i \vec{c}_i, \quad \textrm{s.t.}\quad \vec{d}_j^T\vec{a} = 0 \quad \textrm{for}\quad j\in\{1,\dots,N\},
\end{equation}
where $\mat{A}$ and the $\mat{B}_i$ are $M\times M$ matrices and $M>N$. We can begin by extending the $\vec{d}_j$ to a basis of $\mathrm{I\!R}^M$ by introducing $M-N$ orthonormal vectors $\vec{e}_k$ for $k\in\{1,\dots,M-N\}$ such that each $\vec{e}_k$ is orthogonal to the $\vec{d}_j$. Therefore the vector $\vec{a}$ lies in the $M-N$ dimensional space spanned by the $\vec{e}_k$ and can be written as
\begin{equation}
\vec{a} = \sum_{k = 1}^{M-N} a'_k \vec{e}_k.
\end{equation}
We now define the vector valued function
\begin{equation}
\vec{f}(\vec{x}) = \left[ \mat{A} - \sum_{i = 1}^N \lambda_i \mat{B}_i \right] \vec{a} - \vec{c}_0 - \sum_{i=1}^N \lambda_i \vec{c}_i,
\end{equation}
for $\vec{x} = [\lambda_1,\dots,\lambda_N,a_1',\dots,a_{M-N}']^T$ so solving for our eigenvalues and eigenvectors is equivalent to finding roots of $\vec{f}(\vec{x}) = \vec{0}$. Gradients of $\vec{f}$ with respect to the $\lambda_i$ and $a'_k$ can easily be determined analytically and standard gradient based techniques can be used to find these roots. Alternatively, numerical techniques developed specifically for multi-parameter problems \citep{MPEVP_book,Two_Par_EVP} may prove effective, however, these rely on converting the problem to a large single parameter problem which scales poorly with $N$ and $M$.

\bibliographystyle{jfm}
\bibliography{bibliography}

\begin{thebibliography}{21}
\expandafter\ifx\csname natexlab\endcsname\relax\def\natexlab#1{#1}\fi

\bibitem[Atkinson(1972)]{MPEVP_book}
{\sc Atkinson, F.~V.} 1972 {\em Multiparameter Eigenvalue Problems\/}. Academic Press.

\bibitem[Born \& Wolf(2019)]{BornW19}
{\sc Born, M. \& Wolf, E.} 2019 {\em Principles of Optics\/}, 7th edn. CUP.

\bibitem[Brion {\em et~al.\/}(2014)Brion, Sipp \& Jacquin]{BrionSJ14}
{\sc Brion, V., Sipp, D. \& Jacquin, L.} 2014 Linear dynamics of the {L}amb-{C}haplygin dipole in the two-dimensional limit. {\em Phys. Fluids\/} {\bf 26}, 064103.

\bibitem[Crowe \& Johnson(2023)]{Crowe_Johnson_23}
{\sc Crowe, M.~N. \& Johnson, E.~R.} 2023 The evolution of surface quasi-geostrophic modons on sloping topography. {\em J. Fluid. Mech.\/} {\bf 970}, A10.

\bibitem[Davies {\em et~al.\/}(2023)Davies, Sutyrin, Crowe \& Berloff]{Davies_et_al_2023}
{\sc Davies, J., Sutyrin, G.~G., Crowe, M.~N. \& Berloff, P.~S.} 2023 {Deformation and destruction of north-eastward drifting dipoles}. {\em Phys. Fluids\/} {\bf 35}~(11), 116601.

\bibitem[Flierl \& Haines(1994)]{FlierlH94}
{\sc Flierl, G.~R. \& Haines, K.} 1994 The decay of modons due to {R}ossby wave radiation. {\em Phys. Fluids\/} {\bf 6}, 3487--3497.

\bibitem[Fl\'{o}r {\em et~al.\/}(1995)Fl\'{o}r, van Heijst \& Delfos]{FlorVD95}
{\sc Fl\'{o}r, J.~B., van Heijst, G. J.~F. \& Delfos, R.} 1995 Decay of dipolar vortex structures in a stratified fluid. {\em Phys. Fluids\/} {\bf 7}, 374--383.

\bibitem[Johnson \& Crowe(2021)]{JohnsonC21}
{\sc Johnson, E.~R. \& Crowe, M.~N.} 2021 The decay of a dipolar vortex in a weakly dispersive environment. {\em J Fluid Mech\/} {\bf 917}~(A35).

\bibitem[Johnson \& Crowe(2023)]{Johnson_Crowe_23}
{\sc Johnson, E.~R. \& Crowe, M.~N.} 2023 Oceanic dipoles in a surface quasi-geostrophic model. {\em J. Fluid. Mech.\/} {\bf 958}, R2.

\bibitem[Kizner {\em et~al.\/}(2003)Kizner, Berson, Reznik \& Sutyrin]{Kizner_et_al_2003}
{\sc Kizner, Z., Berson, D., Reznik, G. \& Sutyrin, G.} 2003 The theory of the beta-plane baroclinic topographic modons. {\em Geophys. Astrophys. Fluid Dyn.\/} {\bf 97}~(3), 175--211.

\bibitem[Lamb(1932)]{Lamb_1932}
{\sc Lamb, H.} 1932 {\em Hydrodynamics\/}. Cambridge University Press.

\bibitem[Larichev \& Reznik(1976)]{LarichevR76d}
{\sc Larichev, V.~D. \& Reznik, G.~M.} 1976 Two-dimensional solitary {R}ossby waves. {\em Dokl. Akad. Nauk SSSR\/} {\bf 231}, 12--13.

\bibitem[Makarov \& Kizner(2011)]{makarov_kizner_2011}
{\sc Makarov, V.~G. \& Kizner, Z.} 2011 Stability and evolution of uniform-vorticity dipoles. {\em J. Fluid Mech.\/} {\bf 672}, 307–325.

\bibitem[McWilliams(1980)]{mcwilliams_1980}
{\sc McWilliams, J.C.} 1980 An application of equivalent modons to atmospheric blocking. {\em Dyn. Atmos. Oceans\/} {\bf 5}, 43--66.

\bibitem[Meleshko \& van Heijst(1994)]{MeleshkoH94}
{\sc Meleshko, V.~V. \& van Heijst, G. J.~F.} 1994 On {C}haplygin's investigations of two-dimensional vortex structures in an inviscid fluid. {\em J. Fluid Mech.\/} {\bf 272}, 157--182.

\bibitem[Muhi{\u c} \& Plestenjak(2010)]{Two_Par_EVP}
{\sc Muhi{\u c}, A. \& Plestenjak, B.} 2010 On the quadratic two-parameter eigenvalue problem and its linearization. {\em Linear Algebra and its Applications\/} {\bf 432}~(10), 2529--2542.

\bibitem[Ni {\em et~al.\/}(2020)Ni, Zhai, Wang \& Hughes]{NiZWH20}
{\sc Ni, Q., Zhai, X., Wang, G. \& Hughes, C.~W.} 2020 Widespread mesoscale dipoles in the global ocean. {\em J. Geophys. Res. Oceans\/} {\bf 125}, e2020JC016479.

\bibitem[Nielsen \& Rasmussen(1997)]{NielsenR97}
{\sc Nielsen, A.~H. \& Rasmussen, J.~J} 1997 Formation and temporal evolution of the {L}amb dipole. {\em Phys. Fluids\/} {\bf 9}, 982--991.

\bibitem[Nycander \& Isichenko(1990)]{NycanderI90}
{\sc Nycander, J. \& Isichenko, M.~B.} 1990 Motion of dipole vortices in a weakly inhomogeneous medium and related convective transport. {\em Phys. Fluids\/} {\bf 2}, 2042--2047.

\bibitem[Protas(2024)]{Protas_2024}
{\sc Protas, B.} 2024 On the linear stability of the {L}amb-{C}haplygin dipole. {\em J. Fluid Mech.\/} p. in press.

\bibitem[Rostami \& Zeitlin(2021)]{rostami_zeitlin_2021}
{\sc Rostami, M. \& Zeitlin, V.} 2021 Eastward-moving equatorial modons in moist-convective shallow-water models. {\em Geophys. Astrophys. Fluid Dyn.\/} {\bf 115}~(3), 345--367.

\end{thebibliography}

\end{document}